
%
%
\input harvmac
\pretolerance=10000

\Title{HWS-92/09, hep-th/9301008}{A Simple Method for Computing
Soliton Statistics}


\centerline{Donald Spector\footnote{$^\dagger$}
{spector@hws.bitnet}}
\bigskip\centerline{Department of Physics, Eaton Hall}
\centerline{Hobart and William Smith Colleges}
\centerline{Geneva, NY 14456 USA}

\vskip .3in
\vskip .3in
I propose a simple method
for identifying the statistics of kink-type solitons
in a large class of theories.  The method is based on the Witten index,
but can in fact be used to determine soliton statistics in
non-supersymmetric theories as well.

\Date{11/92; rev. 6/94} 


The spectra of field theories with multiple classical vacua include
kinks, solitons corresponding to field configurations that
interpolate between vacua.
Even though such solitons are field
configurations of a bosonic field, the kinks themselves can
in fact be fermionic \ref\kfermi{D. Finkelstein and J. Rubinstein,
{\it J. Math. Phys.} {\bf 9} (1968) 1762.}.
In this letter, I will propose an extremely simple method for identifying
the statistics of certain kinks.  The method is so simple because
it takes advantage of the Witten index, and the Witten index can be
computed reliably in very simple ways
\ref\trf{E. Witten, {\it Nucl. Phys.} {\bf B202} (1982) 253\semi
S. Cecotti and L. Girardello, {\it Phys. Lett.} {\bf 110B} (1982) 39\semi
R. Akhoury and A. Comtet, {\it Nucl. Phys.} {\bf B246} (1984) 253.}.
Importantly, however, the applicability of this method is not restricted to
supersymmetric theories.  Because a discrete quantity, such as the
statistics, cannot change under continuous deformations of the parameters
of a theory, if we know the statistics of the kink in a supersymmetric
theory, we know too the statistics of the kink in any non-supersymmetric
theory which can be reached by a smooth change in the parameters of the
original theory.  In this way, then, we can use results on statistics
obtained in supersymmetric theories (where the Witten index turns out
to make the relevant calculations simple) to obtain results about
statistics in non-supersymmetric theories.

This paper is meant to present the method and to highlight its
essential features, and so I will proceed in the way that seems
most effective, which is by means of an example, followed by more
general discussion.
The model I consider is in $2+1$
dimensions, but there is nothing special about this case; as will be
readily apparent, the method I use is valid in any number of dimensions
(although, of course, results in specific models will vary).\foot{The
potential role of anyons, which are special to $2+1$ dimensions,
is discussed in the final section of this paper.}

The outline of the paper is as follows.  First, we see how in
supersymmetric theories the Witten index leads to an easy identification
of the statistics of the states corresponding to classical vacua. I then
argue that the determination of vacuum statistics leads to a determination
of the statistics of the kink solitons that interpolate between those vacua.
Next, I discuss the extension of these results to non-supersymmetric
theories by means of continuity arguments, and thus find a large
class of non-supersymmetric theories in which our arguments easily give
the kink statistics.  I then discuss some aspects of the general application
of this method, and present briefly some of the
results and implications of this method in
other theories.  I close with mention of some open problems.

To begin the argument, then,
consider a $2+1$-dimensional model with $N=1$ supersymmetry. (See
\ref\GGRS{S.J. Gates, M.T. Grisaru, M. Ro\v cek, and W. Siegel,
{\it Superspace, or One Thousand and One Lessons in Supersymmetry}
(Benjamin/Cummings, Reading, MA, 1983).}
for a review
of $2+1$ dimensional supersymmetry.) The
simplest such theories may be described in terms of a single
real scalar superfield $\Phi(x^\mu,\theta^\alpha)$,
where $\theta^\alpha$ is a real,
two-component anticommuting coordinate. The superfield $\Phi$
contains as its physical components a real scalar field $\phi$ and
a real, two-component fermion $\psi^\alpha$. Defining the covariant derivative
$D_\alpha = \partial/\partial \theta^\alpha
+ i \theta^\beta \gamma^\mu_{\alpha\beta} \partial_\mu$,
we can write a supersymmetric Lagrangian
\eqn\lags{{\cal L}=\int d^2\theta \bigl(
   {1\over 2}D^\alpha\Phi D_\alpha\Phi + W(\Phi) \bigr)~,}
where $W(\Phi)$ is the superpotential. In terms of the physical
components, this theory has Lagrangian
\eqn\lagc{{\cal L}=
   {1\over 2}\partial^\mu \phi \partial_\mu \phi
  + {i\over 2} {\bar \psi}\gamma^\mu \partial_\mu \psi
  + \bigl({\partial W \over \partial \phi}\bigr)^2
  +{\partial^2 W \over \partial \phi^2}{\bar \psi}\psi ~.}

The Witten index of a supersymmetric theory is the number of bosonic zero
energy states minus the number of fermionic zero energy states,
written formally as $tr(-1)^F$ \trf.
Because states of non-zero energy in a supersymmetric theory must appear
in bose-fermi pairs, the index may be calculated exactly using an
approximation scheme that respects supersymmetry; for example, both
perturbative
and semiclassical arguments give exact values for the Witten index.
Similarly, the index is unaffected by supersymmetry-preserving
changes in the parameters of a theory (as long as these
do not change the behavior of the potential at infinity); states
can only be lifted from or lowered to zero energy in bose-fermi
pairs under such variations, and so such variations cannot change the
Witten index.

We now consider evaluating the index for theories of the type described
above in \lags. For the purposes of this paper, it is sufficient
to restrict our attention to superpotentials of the form
\eqn\spot{W(\Phi) = {g\over 3}\Phi^3 - \alpha \Phi ~,}
where $g$ and $\alpha$ are real. Without loss of generality,
we take $g>0$.
Note that the Witten index in this theory is unaffected by
changes in $\alpha$. We will be interested in comparing the Witten
index at negative and positive values of $\alpha$.

We will calculate the Witten index by identifying the minima of the
classical potential which have zero energy,
and then examining the perturbative spectrum about each such minimum.
Supersymmetric classical minima are given by the condition
\eqn\smin{{\partial W \over \partial \phi} =0 ~,}
which in our case yields
\eqn\ephi{\phi^2 = {\alpha\over g}~.}

In the case that $\alpha$ is negative, this equation \ephi\ has no
solutions, as $\phi$ is a real field. Consequently, at the classical
level already, there are no zero energy states, and so the
Witten index is zero.

In the case that $\alpha$ is positive, however, this equation \ephi\
has two solutions,
\eqn\phisol{\phi = \pm \sqrt{\alpha \over g}~.}
It is easy to check that all the perturbative excitations about each
of these minima are massive. Consequently, the Witten index
is given simply by counting up the classical vacua. Here there are
two such vacua, and so, initially, one might expect the Witten index
to have value two.

However, the Witten index cannot be affected
by changes in $\alpha$, and we saw unambiguously that,
for $\alpha < 0$, the Witten index is zero. Therefore, the
Witten index must be zero even for $\alpha > 0$.  In order
for this to be so, one of the vacua in \phisol\ must be bosonic
and the other
vacuum must be fermionic. Then these two classical vacua
enter the Witten index sum with opposite signs,
producing a Witten index of zero.

Thus we have established quite easily that the two vacua have opposite
fermion number. Let us now consider the theory with $\alpha >0$
a little bit further. As we know from \phisol, the scalar potential
\eqn\scpot{V(\phi) = \bigl(g\phi^2 - \alpha\bigr)^2~,}
has two minima. There are kink-type
solitons which interpolate between these two classical vacua.
What can we say regarding the statistics of such solitons?
In the next paragraphs, I would like to argue that the only physically
sensible picture is that these solitons are themselves fermions.

The physical idea underlying our argument is that since
we can move from one minimum of the potential to the other by,
loosely speaking, acting on the initial vacuum state with a
soliton operator, and since the vacua have opposite fermion
number, then the soliton itself must be a fermion, not a boson.
We will now refine this informal argument into a
more fully developed physical picture.

Let the two minima of the potential occur at $\phi(x^\mu)=\phi_1$ and
$\phi(x^\mu)=\phi_2$.
Suppose we have an operator that converts the first vacuum into the second
one; clearly such an operator must be fermionic.  Suppose, then, that
a soliton is created, passes through space, and continues on far past
some oberver.  What does such an observer see?  Initially, the observer
sees the first vacuum state, $\phi(x^\mu)=\phi_1$.
The soliton is created (i.e., a soliton
creation operator acts on this vacuum), so that
the field configuration $\phi(x^\mu)$ is an interpolation from
$\phi_2$ to $\phi_1$. This soliton then goes past the observer.
After a sufficiently long time, the observer would see
the state of the system to be, to arbitrarily
good accuracy, indistinguishable from the second vacuum, for which
$\phi(x^\mu)=\phi_2$ everywhere.  If no soliton
had appeared, this transition would not have occurred.  Thus, the
action of the soliton on the first vacuum has been, in effect, to convert
it into the second vacuum.  Evidently, then, the kink soliton must be
fermionic.

We can refine this argument further by placing space on a tube with
the topology of ${R}\times S^1$.
To have an easy way to refer to
the ends of this tube, let the {\it left} end be the end of the tube at
$-\infty$ on the $R$-axis, and let the {\it right} end be the end of the
tube at $+\infty$ on the $R$-axis. With $\phi_1$ and $\phi_2$ the field
values corresponding to the two classical vacua, one ordinarily
formulates this theory with four sectors, corresponding respectively
to the following boundary conditions:
(1) $\phi = \phi_1$ at both ends;
(2) $\phi = \phi_2$ at both ends;
(3) $\phi = \phi_2$ on the left and $\phi =\phi_1$ on the right;
and (4) $\phi = \phi_1$ on the left and $\phi = \phi_2$ on the right.
The lowest energy state in each of these sectors corresponds
respectively to the first classical vacuum; the second classical vacuum; a
soliton; and an anti-soliton.

However, for our purposes, it is advantageous to
use an alternative formulation of the theory in which there
are only two sectors.  These sectors are given, respectively,
by the following boundary conditions: (1) $\phi = \phi_1$ on the
left; and (2) $\phi = \phi_2$ on the left.  The lowest energy states
in these
two sectors are, respectively, the first and the second classical vacua,
since in each sector, the energy is minimized by $\phi(x^\mu)$
having a uniform value, with the particular value fixed by the boundary
condition on the left.
Thus we still have the same vacuum structure as with the more familiar
formulation.\foot{Note, too, that for an infinitely long tube, there
is no tunneling between these vacua.}
The kink field configurations still exist with the boundary conditions
of the two-sector formulation;
they simply
are not stable.  The instability of the soliton is not a problem here; it is,
in fact, an advantage, as we can determine some of the properties of
the soliton by studying what it can evolve into,
just as we routinely determine the properties of particles
produced at accelerators by studying their decays.

Given a kink field configuration, which interpolates between
the two classical vacua, it can be deformed continuously
into a vacuum configuration, while staying within the space of
field configurations which have finite energy and respect the
boundary condition, through a continuous change
in the value of the field at the {\it right} end of the tube.
(The end of the tube is a circle of finite radius.)
Consequently, to an observer near the right end of the
tube, the soliton will not appear as a sensible approximate notion.
However, to an observer near the {\it left} end of the tube, as long
as the tube is extremely long, the soliton
{\it will} appear as a sensible approximate notion.  Thus, near the
left end of the tube, it is
physically sensible to
speak of creating a kink, and then to watch its
evolution in time.  We can use the quantum numbers of the final state to
help us determine the quantum numbers of the soliton.

So now let us consider this two-sector version of the theory, and
let us imagine
that initially the system is in the first vacuum state.  Now imagine
a soliton is created at the left end of the tube, so that
$\phi=\phi_2$ on the left, but $\phi=\phi_1$ still on the right,
with the velocity of the
soliton pointing toward the right end of the tube.  As the kink
moves rightward, it leaves behind a longer and longer stretch over which
the field $\phi$ has the value $\phi_2$.  In the far future, when the
disturbance reaches the right end of the tube, the value of the field
at the right end will change smoothly from $\phi_1$ to $\phi_2$ (the
finite radius of the circle $S^1$ makes this possible), leaving the
system finally in the second vacuum state.

In other words, the state created by acting on the first vacuum with
a soliton creation operator evolves into the second vacuum state.  Since
the first and second vacuum states have opposite fermion number, then,
the soliton must itself be fermionic.

Obviously, this all works, as it must, in the other direction, too.
The state created by acting on the second vacuum with
an anti-soliton creation
operator can evolve into the first vacuum, verifying that the
anti-soliton, like the soliton, has fermion number $+1$.

Thus we see that we have used a Witten index calculation to identify
the fermion number of a soliton.
It is striking to have such a
simple way to determine that a soliton is a fermion.
In the remainder of this paper, I examine some of the details
associated with this method, and explore some of its consequences.
These remarks are meant to cover the essentials, not to
be exhaustive.

First, there is a large class of non-supersymmetric theories in
which the above type of argument still determines the statistics
of kink solitons. This is because the statistics of a soliton
will not be changed by any deformation of the parameters of the
theory, including deformations that violate
supersymmetry, as discrete quantum numbers cannot change under
continuous changes in the parameters. Thus in any model which
can be obtained from a supersymmetric theory by continuous
(non-supersymmetric)
changes in the parameters of that theory, and which has essentially
the same structure of classical vacua (e.g., under the change
of parameters, classical vacua should not appear, disappear,
or coalesce), the statistics of the
soliton will be the same as in the supersymmetric theory.

Some clarification of these remarks is in order in the case of
two spatial dimensions, where anyon statistics can occur.
In this paper, we should take the following as the definition of $(-1)^F$
in the $2+1$ dimensional example we study.
The operator $(-1)^F$ is the operator that squares to unity;
anticommutes with the supercharges; and (anti)commutes with those
fundamental fields which are quantized according to
(anti)commutation relations. Since the theories I am considering
can always be formulated in terms of such quantum fields, this
provides a complete definition of $(-1)^F$.
Statistics then refers to the eigenvalue of this operator, which
must be $\pm 1$. In the $2+1$ dimensional model in question,
we have an extended object whose $(-1)^F$ eigenvalue is opposite to
that of the local field out of which it is built.

In three or more spatial dimensions,
such a definition of statistics is identical
to the notions of statistics more conventionally defined.  In
two spatial dimensions, however, there is an alternative
definition of statistics which is more familiar and conventional,
namely statistics defined in terms of the relative phases associated with
multiparticle configurations in the functional integral, which
gives rise to the possibility of fractional statistics realized by
anyons. This definition of statistics is distinct
from the one considered above, and allows,
in 2+1 dimensions, for the possibility of statistics which change
continuously as the parameters of the theory change.  In higher
dimensions, where anyons are not possible, this possibility does not
arise.  In order to preserve a discrete notion of statistics in $2+1$
dimensions, therefore, we use the less conventional definition of
statistics and $(-1)^F$ given above.
It is worth commenting here that in the
example model in this paper, one can argue that the soliton must have
half-integral spin, by using a Witten index
argument based on the conventional
definition of statistics, along with the results of \ref\sany{Z. Hlousek
and D. Spector, {\it Nucl. Phys.} {\bf B344} (1990) 763.}.  Indeed,
one can combine the method here with the results
in \ref\wanyon{D. Spector,{\it Anyon Statistics and the Witten Index},
HWS-94/14.} (which connects these
different notions of statistics) to obtain further results
on kinks in $2+1$ dimensional supersymmetric theories.

Second, one might worry that non-zero fermion number might be emitted
in the evolution of a soliton into a vacuum state.  We can use
the invariance of the kink statistics under changes in the parameters
of a theory to argue that this problem does not arise.  Let us
imagine deforming our original theory so that the scalar vacua remain
unchanged, but so that the perturbative fermionic particles are more
massive than the soliton.  This deformation cannot change a discrete
quantum number such as the kink statistics.  It also does not alter our
argument connecting vacuum statistics to kink statistics.
Consequently, in this modification of the original theory,
clearly the only  particles that can carry off the
released energy when the soliton decays are bosons,
and so no statistics are lost to particle
emission.  Thus, we can safely conlude, as we did above,
that the kink is fermionic in
the deformed theory.  By continuity, the soliton must be fermionic
in the original theory as well,
and so even if the perturbative fermions are light enough to be emitted
in the transition from a soliton to a vacuum background,
they must be admitted in pairs to maintain consistency with the
results we obtained at large fermion mass.

Third, note that the basic structure of the argument presented here
does not depend on dimension (despite the special issues associated
with anyons in two spatial dimensions). However, the argument does distinguish
between the cases of real and complex scalar fields. Now in some
dimensions and some situations,
supersymmetry only requires real fields, whereas in
other cases complex fields are necessary. Thus, while
the methods described above may be applied in any number of dimensions,
the conclusions one draws will be different depending upon whether
the supersymmetric theory involves real or complex fields. In fact,
the distinctions between real and complex fields may actually be used
to relate the computation of the Witten index to the fundamental
theorem of algebra (and to the failure of such a theorem to hold
for polynomials over the reals) \ref\fta{D. Spector, {\it Supersymmetry,
Vacuum Statistics, and the Fundamental Theorem of Algebra}, HWS-94/15.}.

Fourth, it is worth noting that in such a familiar example as the
sine-Gordon model, our argument reproduces in simple fashion the
familiar result that the solitons in that model are fermions \fta .

An important extension of this argument would be to generalize it to
determine the statistics of other extended objects which do not have
the same statistics as their constituent field(s). Topological solitons
such as the skyrmion provide a prime example of this transmutation
of statistics, yet are not covered by the argument presented above.

Finally, it would be instructive to connect the arguments of this paper
with more conventional treatments of soliton statistics. For example,
one can directly determine the statistics of classical vacua; indeed,
performing such a calculation in supersymmetric quantum
mechanics explicitly yields a structure of the kind inferred
via Witten index arguments in the
above model \fta.
One can already see from
the results presented that there is a kind of ``Berry's sign''
that arises in mapping from one of the vacua to the other in the
model theory I have discussed in this paper.

\bigbreak\bigskip\bigskip\centerline{{\bf Acknowledgments}}\nobreak
I thank B. Greene and A. Shapere for conversations.
I also thank the LNS Theory Group at Cornell University, where
parts of this research were conducted, for its hospitality.
This work was
supported in part by NSF Grant. No. PHY-9207859 and by a Hobart and
William Smith Faculty Research Grant.

\listrefs
\bye